\documentclass[aps,prc,amsmath,amssymb,twocolumn,showpacs]{revtex4-1} 
\usepackage{graphicx,amsmath,amssymb,bm,soul}

\usepackage[usenames]{color}

\begin{document} 
 
 \title{\textit{Ab initio} and phenomenological studies of the static response of neutron matter}

\author{Mateusz Buraczynski and Alexandros Gezerlis}
\affiliation{Department of Physics, University of Guelph, 
Guelph, ON N1G 2W1, Canada}

\begin{abstract} 
We investigate the problem of periodically modulated strongly interacting neutron matter. We carry out \textit{ab initio} non-perturbative auxiliary-field diffusion Monte Carlo calculations using an external sinusoidal potential in addition to 
phenomenological two- and three-nucleon interactions. Several choices for the wave function ansatz are explored and special care is taken to extrapolate finite-sized results to the thermodynamic limit. We perform calculations at various densities as well as at different strengths and periodicities of the one-body potential. Our microscopic results are then used to constrain the isovector term from energy-density functional theories of nuclei at many different densities, while making sure to separate isovector contributions from bulk properties. Lastly, we use our results to extract the static density-density linear response function of neutron matter at different densities. Our findings provide insights into inhomogeneous neutron matter and are related to the physics of neutron-star crusts and neutron-rich nuclei.
\end{abstract} 
\maketitle 
\section{Introduction}
Neutron matter is integral to the study of neutron stars and neutron-rich nuclei \cite{Gandolfi:2015}. The equation of state (EOS) of neutron matter has been studied extensively using \textit{ab initio} approaches~\cite{Friedman:1981,Akmal:1998,Schwenk:2005,Gezerlis:2008,Epelbaum:2008b,Kaiser:2012}. While neutron matter does not obtain in nature in pure form, its EOS is closely connected to that of physical systems. More specifically, it is direct input into the Einstein's field equations (typically cast as
the Tolman-Oppenheimer-Volkov equations) that lead to basic observable properties of a neutron star. 
On the other hand, the neutron-matter EOS is also connected to nuclei via the use of nuclear energy density functionals (EDFs).
EDFs take on a variety of forms ~\cite{Bender:2003,SLy,RiosGogny,Lacroix} and are typically fit to empirical data such as nuclear masses and radii. Other constraints may include the EOS of neutron matter~\cite{Fayans,SLy,Brown:2000, Gogny,Fattoyev,Brown:2014,Rrapaj:2015}, the neutron-matter pairing gap~\cite{Chamel:2008}, the energy of a neutron impurity~\cite{Forbes:2014,Roggero:2014b}, or the properties of a neutron drop~\cite{Pudliner:1996,
Pederiva:2004,Gandolfi:2011,Potter:2014,Gandolfi:2016}. The EDF approach is especially useful as it allows for predictions to be made across the nuclear chart. Investigating neutron matter is an excellent opportunity for benchmarking both phenomenological~\cite{Carlson:Morales:2003,Gandolfi:2009,Gezerlis:2010,Gandolfi:2012,Baldo:2012} 
and chiral~\cite{Hebeler:2010,Tews:2013,Gezerlis:2013,Coraggio:2013,Hagen:2014,Gezerlis:2014,Carbone:2014,Roggero:2014,Wlazlowski:2014,Tews:2016} nuclear interactions and many-body methods. 

While homogeneous matter is an intriguing system, it is not fully representative of either finite systems or astrophysical 
settings, since nuclei and neutron-star matter are inhomogeneous systems. 
The matter in a neutron-star crust is rich with unbound neutrons and also contains a lattice of nuclei. Focussing only
on the unbound neutrons, these experience the interaction with the lattice as a periodic modulation. Similarly, nuclei
are finite systems: their density eventually drops off as one goes farther away from the center of the nucleus. Thus, 
as a first approximation, one can study the effects of a one-body external periodic potential on pure infinite neutron matter.
This system, periodically modulated neutron matter, directly mimics the situation in a neutron star; also, if used as an input constraint to EDFs, it can inform us about the physics of neutron-rich nuclei.
This problem of an external periodic modulation is known as the \textit{static response} of neutron matter: it amounts to 
a comparison between externally modulated and unmodulated infinite neutron systems.
This problem has been tackled using a variety of approximations in the literature~\cite{Iwamoto:1982,Olsson:2004,Chamel:2011,Chamel:2013,Kobyakov:2013,Pastoretheory,Chamel:2014,Davesne:2015}, see Ref.~\cite{PastoreReport} for a recent review.
The static response problem has also received a lot of attention in other areas of physics~\cite{Pines:1966}.
This includes pioneering quantum Monte Carlo calculations for strongly correlated systems such as liquid $\rm {{}^4He}$ at zero temperature and pressure \cite{Moroni:1992} and the three-dimensional electron gas \cite{Moroni:1995}. 

In a recent Letter~\cite{Buraczynski&Gezerlis:2016} we reported on the first \textit{ab initio} calculation 
of the static response of 
neutron matter. This included a calculation of the linear density-density static response function of neutron matter following from a Quantum Monte Carlo (QMC) $T=0$ approach. Specifically, this involved the Variational Monte Carlo (VMC) and auxiliary-field diffusion Monte Carlo (AFDMC) methods: these are non-perturbative and accurate methods for computing properties of many-body nuclear systems. Reference~\cite{Buraczynski&Gezerlis:2016} also mentioned the role of 
finite-size effects, the wave-function
ansatz, and conclusions that can be drawn on the isovector gradient coefficient of EDFs. 

In the present article, 
we expand on Ref.~\cite{Buraczynski&Gezerlis:2016}, by discussing finite-size effects and the wave-function ansatz 
in more detail. In addition to this, we have carried out new QMC calculations for the periodically modulated system
at many different densities, extracting from there EDF parameters for several
different Skyrme parametrizations. We have also carried out several new calculations
of the modulated system for several different strengths and periodicities of the external one-body potential: this includes
a study of the lower-density regime, as well as separate simulations (at both low and intermediate densities) with and 
without three-nucleon interactions. These results provide a more detailed understanding of different effects and, since
they are non-perturbative and accurate, they provide ``synthetic data'' that can be used in (or compared to) other approaches.
We make some first steps in this direction by examining the response function coming from selected EDFs.

We begin with some background on the methods employed (section~\ref{sec:methods}), mainly establishing our notation 
in what follows. We also provide a more extensive discussion of the theory of static response (section~\ref{sec:static}), 
explaining what the density-density response is, and how it can be extracted from our calculations. 
We then proceed to study finite-size
effects in the non-interacting problem 
in detail (section~\ref{sec:finite}), before turning to the QMC and EDF results for the interacting and periodically
modulated problem (section~\ref{sec:results}). 

\section{Methods}
\label{sec:methods}
\subsection{Auxiliary-field diffusion Monte Carlo}

The Hamiltonian of the full interacting problem we study is made up of a non-relativistic kinetic term, a two-nucleon (NN) interaction, a three-nucleon (NNN) interaction, and a one body potential:

\begin{equation}
\hat{H}=-\frac{\hbar^2}{2m}\sum_{i}{\nabla_i^2}+\sum_{i<j}{v_{ij}}+\sum_{i<j<k}{v_{ijk}}+v_{\rm ext},
\label{eq:I_H}
\end{equation}
where $v_{\rm ext}=\sum_{i}{v(\mathbf{r}_i)}$ and $v(\mathbf{r}_i)=2v_{q}\cos(\mathbf{q}\cdot \mathbf{r}_i)$. The external potential is motivated by the periodicity of nuclei found in a neutron-star crust. 

Diffusion Monte Carlo (DMC) is an accurate method for computing the ground-state energy 
of a many-body system. The procedure takes a trial wave function $|\Psi_T \rangle$ as input and projects 
the ground-state out by evolving forward in imaginary time \cite{Pudliner:1997}:

\begin{equation}
|\Phi_0 \rangle= \lim_{\tau \to \infty} |\Phi(\tau)\rangle=\lim_{\tau \to \infty} e^{-(\widehat{H}-E_T)\tau}|\Phi(0)\rangle,
\label{eq:prop}
\end{equation}
where $|\Phi(0)\rangle=|\Psi_T\rangle$ and $E_T$ is an energy offset. We handle the fermion sign problem
with the fixed-node (or, more generally, constrained-path) approximation. As a result the method is only exact when the nodal structure of 
$\langle{\bf R}|\Psi_T\rangle$ is the same as that of the ground-state. The trial wave function (including spins) 
takes the form:
 
 \begin{equation}
|\Psi_T \rangle =\prod_{i<j}f(r_{ij})\,\,\mathcal{A}\bigg[\prod_i|\phi_i,s_i\rangle\bigg]
\label{eq:trial}
\end{equation}
 
 This is a product of a Jastrow factor and a Slater determinant of single-particle orbitals. The Jastrow factor has no 0s so it does not impact the nodal structure of the many-body trial wave function. It serves to make the algorithm more efficient since we run simulations for a finite amount of time.
 We use variational Monte Carlo (VMC) to optimize $\langle{\bf R}|\Psi_T\rangle$ by minimizing $\langle \Psi_T|\widehat{H}|\Psi_T\rangle$ over some set of variational parameters. This also 
 serves to produce a set of configurations that sample $|\langle{\bf R}|\Psi_T\rangle|^2$ using the Metropolis algorithm. 
 The propagation in imaginary time is done in small time steps $\Delta \tau$. The Trotter-Suzuki 
 approximation allows the use of Green's functions to carry out the evolution of the wave function. The procedure is a diffusive process that adjusts the configurations by sampling in coordinate space and taking
  into account the potential energy. The actual sampling that is carried out in practice is called importance sampling.
We sample $\langle{\bf R}|\Phi(\tau)\rangle\langle{\bf R}|\Psi_T\rangle$ which has several advantages over 
sampling $\langle{\bf R}|\Phi(\tau)\rangle$ and allows the use of a 
 mixed estimator in extracting the ground-state energy. The mixed estimator is given by:
 \begin{equation}
 E_0=\lim_{\tau \to \infty}\frac{\langle \Phi(\tau)|\widehat{H}|\Psi_T\rangle}{\langle \Phi(\tau)|\Psi_T\rangle}
 =\frac{1}{M}\sum_ME_L({\bf R}_M),
\label{eq:estimator}
\end{equation}
 where $E_L({\bf R})=\langle{\bf R}|\widehat{H}|\Psi_T\rangle/\langle{\bf R}|\Psi_T\rangle$ is called the 
 local energy and is averaged over the configurations. 

 Auxiliary Field Diffusion Monte Carlo (AFDMC) is an extension of DMC that is employed for Hamiltonians with a complicated spin dependence. The method uses a technique that reduces the number of operations for handling 
 spin from exponential to linear. This is achieved by sampling over a set of auxiliary fields that evolve the spin
 component of the wave function \cite{Schmidt:1999}. The number of operations in AFDMC scales with $N^3$. This realistically limits us to simulations on the order of 100 particles.

\subsection{Energy-density functionals}

Another method for computing the energy of this system comes from density functional theory (DFT). In this case,
the nuclear force takes a phenomenological form of the Skyrme type. The many-body wave function takes the form of a Slater determinant of single-particle orbitals $\psi_i(\mathbf{r})$. The ground-state energy is given by:
\begin{equation}
E=\int \mathcal{H}(\mathbf{r})d^3r,
\label{eq:Eint}
\end{equation}
where $\mathcal{H}$ is the energy density energy functional (EDF) \cite{Bender:2003}:
\begin{align}
\mathcal{H} &= \frac{\hbar^2}{2m}\tau+2v_q\cos(\mathbf{q}\cdot \mathbf{r}) n+{\cal E}_{Sk}
\label{eq:Edensity}
\end{align}
and
\begin{align}
&n(\mathbf{r})=\sum_i[\psi_i(\mathbf{r})]^2 \nonumber \\
&\tau(\mathbf{r})=\sum_i[\nabla\psi_i(\mathbf{r})]^2
\label{eq:rho}
\end{align}
are the nucleon and kinetic energy densities. The Skyrme interaction terms of the EDF in the isospin representation are:
\begin{equation}
{\cal E}_{Sk} = \sum_{T=0,1} \left [ ( C^{n,a}_T + C^{n,b}_T n^{\sigma}_0  ) n^2_T 
 + C^{\Delta n}_T (\nabla n_T)^2 + C^{\tau}_T n_T \tau_T  \right ]
\label{eq:Skyrme}
\end{equation}
For pure neutron matter $n_0=n_1=n$ and the same is true for $\tau$. We have done calculations for the SLy4, SLy7, and SkM* parametrizations. $C^{\Delta n}_1$ is known as the isovector gradient term. $C^{\Delta n}_1= -16, -6, \rm{and}\,-17\,\rm{MeV \,fm^5}$ in SLy4, SLy7, and SkM* respectively~\cite{SLy}. In what follows, we will adjust this parameter based on our AFDMC results. Eq.~(\ref{eq:Eint}) provides us with a method to approximate the energy using the density functions. This is called the local-density approximation. Our method of approximating the energy is similar to the VMC optimization of the trial wave function. The $\psi_i$'s are the orbitals of the non-interacting Hamiltonian with our external potential (Mathieu functions). We minimize the right hand side of Eq.~(\ref{eq:Eint}) to get the local-density approximation energies.

\subsection{Static-response theory}
\label{sec:static}

An objective of this work is to compute the linear density-density static response function of neutron matter. This gives a quantitative (up to first order) description of the effect of an external perturbation on the physical properties of a homogeneous neutron gas. Let $\hat{H}_0$ denote the unperturbed Hamiltonian. This is Eq.~(\ref{eq:I_H}) without the $v_{\rm{ext}}=\sum_iv(\mathbf{r}_i)$ term. The $0$ subscript is our notation for the unperturbed system. (Note that this is different from the, standard, $n_0$ notation used in Eq.~(\ref{eq:Skyrme}) in the isospin representation of the EDF.) The ground-state density of the system is a functional of the external potential: $n_v(\mathbf{r})=n_0(\mathbf{r},[v])$ The density-density response functions are defined as the functional derivatives of density with respect to $v$~\cite{Senatore:1999}:
\begin{align}
&n_v(\mathbf{r}) = n_0 + \nonumber \\ 
&\sum_{k=1}^\infty \frac{1}{k!} \int d^3 r_1 \ldots d^3 r_k~\chi^{(k)}(\mathbf{r}_1 - \mathbf{r},\ldots,\mathbf{r}_k - \mathbf{r}) 
v(\mathbf{r}_1)\ldots v(\mathbf{r}_k),
\label{eq:densityexpansion}
\end{align}
where the $\chi^{(k)}$'s are the response functions. $\chi^{(1)}(\mathbf{r})$ is the linear density-density response function. Likewise, the ground state energy can also be expressed as a functional of $v$: $E_v=E_0([v])$. The energy can be expressed as:
\begin{align}
E_v=E_0+\int_0^1d \lambda\int d^3 r n_0({\mathbf{r},[\lambda v]})v(\mathbf{r})
\label{eq:energyPT}
\end{align}
This follows from first-order perturbation theory: $\delta E_v/\delta v(\mathbf{r})=n_v(\mathbf{r})$. This is easy
to see if the interaction term is cast as $\int d^3 r ~\hat{n}(\mathbf{r}) v(\mathbf{r})$, where
the one-body density operator is $\hat{n} = \sum_i \delta(\mathbf{r}-\mathbf{r}_i)$.

The energy and density can be expressed in terms of the Fourier components of the potential $v(\mathbf{r})=\sum_{\mathbf{q}} v_{\mathbf{q}} \exp(i \mathbf{q} \cdot \mathbf{r})$ and the Fourier transforms of the response functions with respect to their spatial arguments $\chi^{(k)}(\mathbf{q}_1,\ldots,\mathbf{q}_k)$:

\begin{align}
&n_v(\mathbf{r})=n_0+\sum_{k=1}^\infty\frac{1}{k!}\sum_{\mathbf{q}_1,\ldots,\mathbf{q}_k} \chi^{(k)}(\mathbf{q}_1,\ldots,\mathbf{q}_k) v_{\mathbf{q}_1}\ldots v_{\mathbf{q}_k} \nonumber \\
&\times \exp[i(\mathbf{q}_1+\ldots+\mathbf{q}_k)\cdot \mathbf{r}] \nonumber \\
&\frac{E_v}{N}=\frac{E_0}{N}+v_0+\frac{1}{n_0} \times \nonumber \\
&\times \sum_{k=1}^\infty\frac{1}{(k+1)!}\sum_{\mathbf{q}+\mathbf{q}_1+\ldots+\mathbf{q}_k=0}\chi^{(k)}(\mathbf{q}_1,\ldots,\mathbf{q}_k)v_{\mathbf{q}}v_{\mathbf{q}_1}\ldots v_{\mathbf{q}_k}
\label{eq:energydensityeqns}
\end{align}

For the one-body external potential $v(\mathbf{r})=2v_{q}cos(\mathbf{q}\cdot \mathbf{r})$ the density is given by:
\begin{align}
&n_v(\mathbf{r})=n_0+2n_{\mathbf{q}} \cos(\mathbf{q}\cdot \mathbf{r}) \nonumber \\
&n_{\mathbf{q}}=\chi^1(q)v_q+\frac{\chi^3(\mathbf{q},\mathbf{q},-\mathbf{q})}{2}v_q^3+\ldots
\label{eq:density}
\end{align}

The change in the density $n_v(\mathbf{r})-n_0$ depends only on odd powers of $v_q$. The change in energy is:
\begin{align}
\frac{E_v}{N}=\frac{E_0}{N}+\frac{\chi^1(q)}{n_0}v_q^2+\frac{\chi^3(\mathbf{q},\mathbf{q},-\mathbf{q})}{4n_0}v_q^4+\ldots
\label{eq:energy}
\end{align}
The energy change only depends on even powers of $v_q$. If the energy per particle is known at several different values of $v_q$ then Eq.~(\ref{eq:energy}) gives a method to calculate lower-order response functions by fitting to a polynomial of even powers. The coefficient of the quadratic term gives the linear density-density response function. The coefficient of the quartic term is very small in our calculations, on the order of $10^{-4}$ MeV$^{-3}$ or smaller. 
Higher-order fits, with more points, are required to reliably extract the third-order response function.

The response of a non-interacting Fermi gas can be computed analytically. It is given by the Lindhard function \cite{Kittel:1969}: 
\begin{equation}
\chi_{L} = - \frac{m  q_F}{2\pi^2 \hbar^2} \left [ 1 + \frac{q_F}{q} 
\left ( 1 - \left ( \frac{q}{2q_F} \right )^2 \right ) \ln \left | \frac{q + 2q_F}{q - 2q_F} \right |   \right ]
\label{eq:Lindhard}
\end{equation}
We compare our results in later sections with this response.

\section{Non-interacting problem}
\label{sec:finite}

Many of the concepts needed for an understanding of interacting particles 
require an understanding of a much simpler problem: the 3D non-interacting free-Fermi 
gas. 
For a finite system of $N$ particles it is standard to restrict positions to a cubic box of 
volume $V=L^3$ and impose periodic boundary conditions on the wave function, when an extended system is the 
end goal of the study. At $T=0$ particles 
occupy states corresponding to the lowest available energy levels. A state is identified by its 
momentum wave-vector ${\bf k}=(2\pi/L) (n_x,n_y,n_z)$ where $n_x$, $n_y$, $n_z$ are
 integers. For a spin-1/2 system a maximum of two particles with different spin-projection can 
occupy the same wave vector, due to the Pauli exclusion principle. A particle placed in the state with 
wave-vector ${\bf k}$ has energy $E=\hbar^2{\bf k}^2/2m$ and occupies the single particle 
orbital $\psi_{\bf k}({\bf r})=e^{i{\bf k}\cdot{\bf r}}/\sqrt V$. Closed shell configurations 
where the energy level populations are filled to capacity occur at $N=\{2,14,38,54,66,114...\}$.
\begin{figure}[t]
\begin{center}
\includegraphics[width=1.0\columnwidth,clip=]{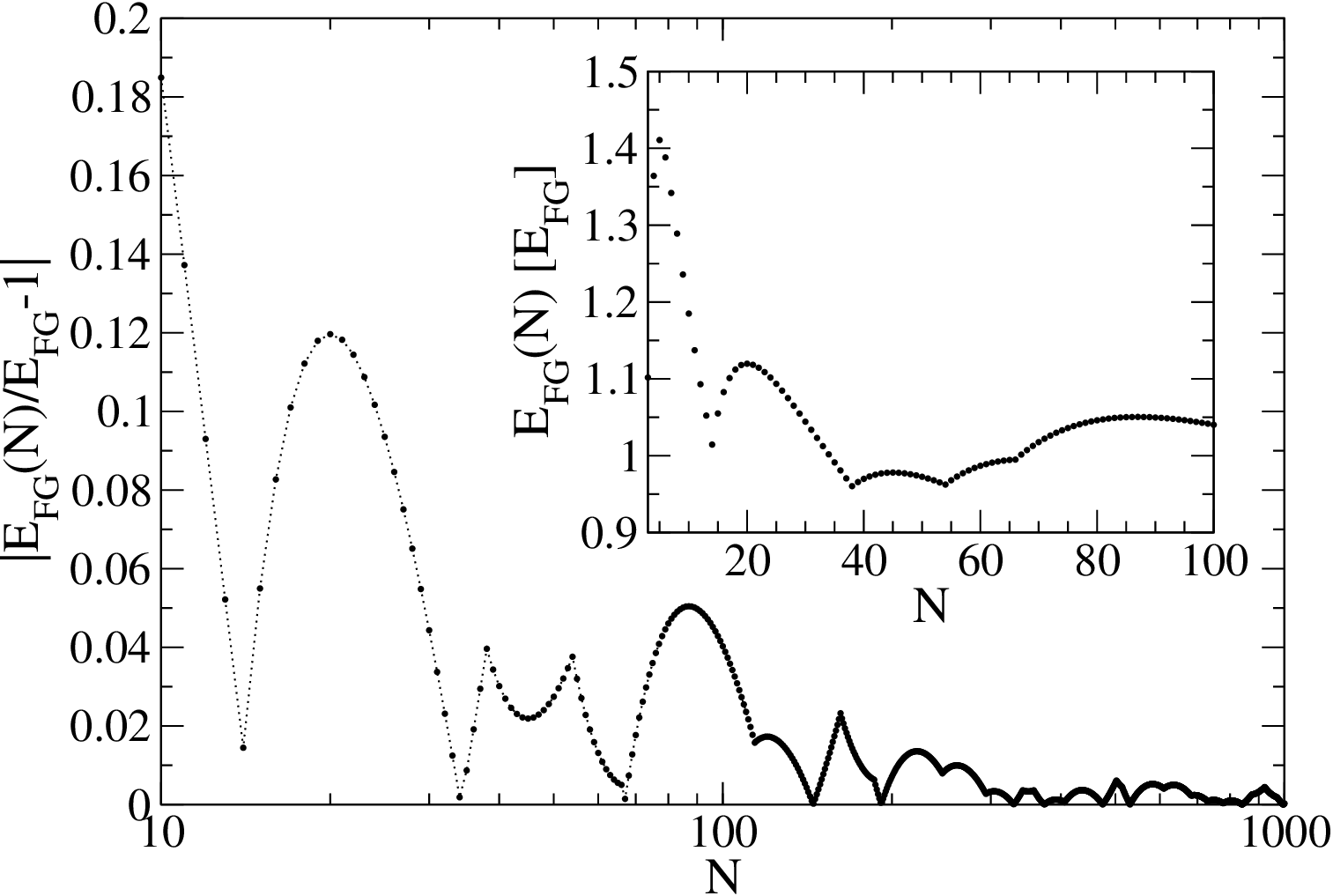}
\caption{FS dependence of the non-interacting free-Fermi gas. FS effects go to zero in the thermodynamic limit. 
The minimum at $N=67$ motivates a study at the closed shell $N=66$. The inset plots the same relationship
 on a linear scale.\label{fig:free}}
\end{center}
\end{figure}
It is preferable to work with closed shells to avoid any ambiguity in determining which
 states are occupied.

Since a neutron star is a macroscopic system we are particularly interested in the
thermodynamic limit (TL) where $N \rightarrow \infty$, $V\rightarrow\ \infty$ and $n = N/V$ is 
constant. In the TL the number density $n$ is related to $k_F$, the maximal wave vector 
magnitude by $k_F=(3\pi^2n)^{1/3}$. The energy per particle is given by $E_{FG}=(3/5)\hbar^2k_F^2/2m
=(3/5)E_F$. Differences in properties between the thermodynamic limit and a finite number of particles
are called finite-size (FS) effects. FS effects go to zero in the thermodynamic limit. They are also largest at small $N$. This can be seen in Fig.~\ref{fig:free} which plots FS effects in the energy per particle versus particle number for the non-interacting free-Fermi gas. This result is density-independent. As mentioned earlier, we are primarily interested in shell closures. 
 \begin{figure}[t]
\begin{center}
\includegraphics[width=1.0\columnwidth,clip=]{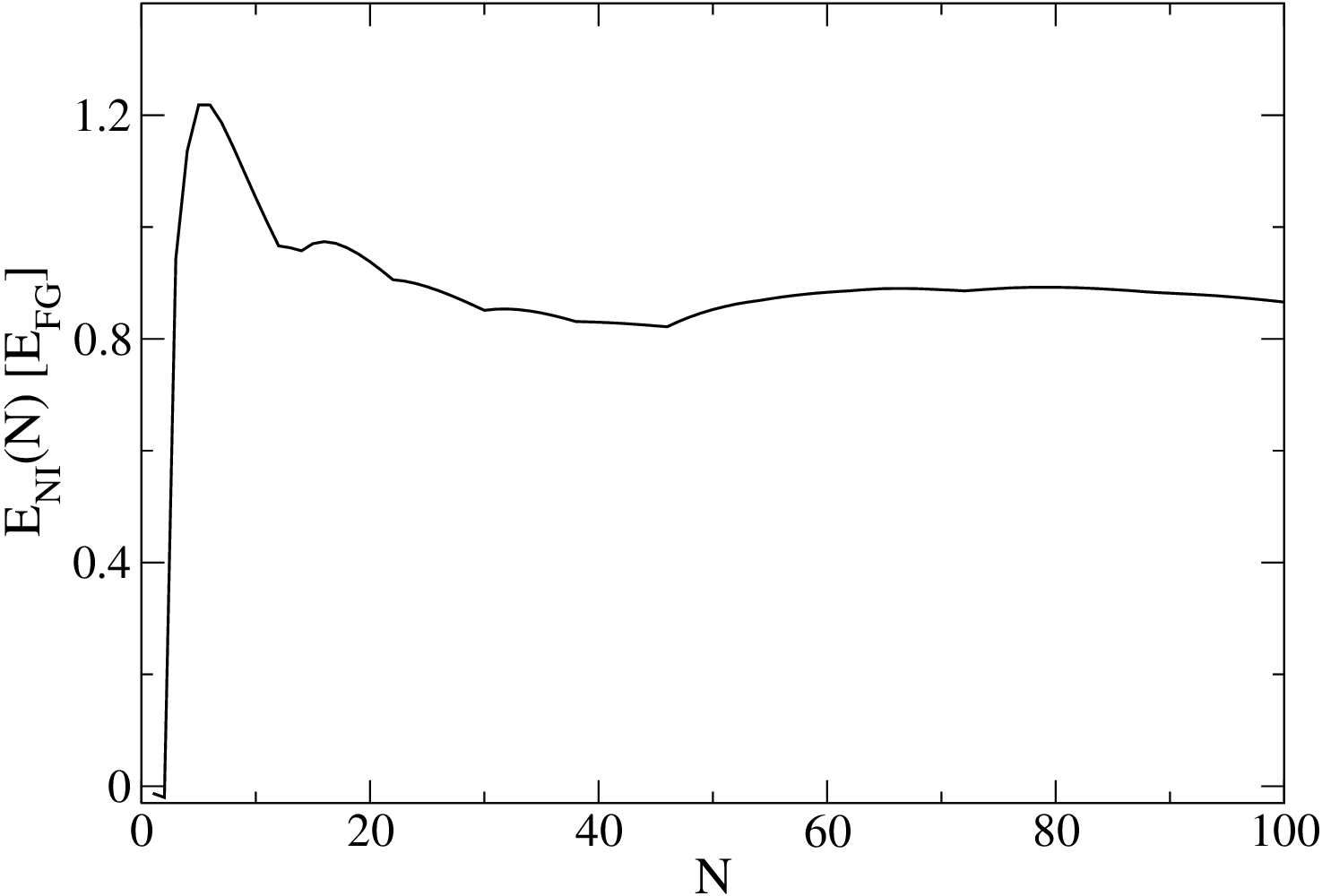}
\caption{FS dependence of the non-interacting Fermi gas in the presence of a one-body potential
of fixed strength $2v_q=0.5E_F$ and periodicity $q=4\pi/L$. $n$ is fixed at $0.1 \,{\rm fm^{-3}}$.\label{fig:NI}}
\end{center}
\end{figure}
These appear at the cusps in the inset of Fig.~\ref{fig:free}. There is a minimum in FS effects occurring at 67 particles. This leads us to study the closed shell at 66 particles.
\begin{figure}[b]
\begin{center}
\includegraphics[width=1.0\columnwidth,clip=]{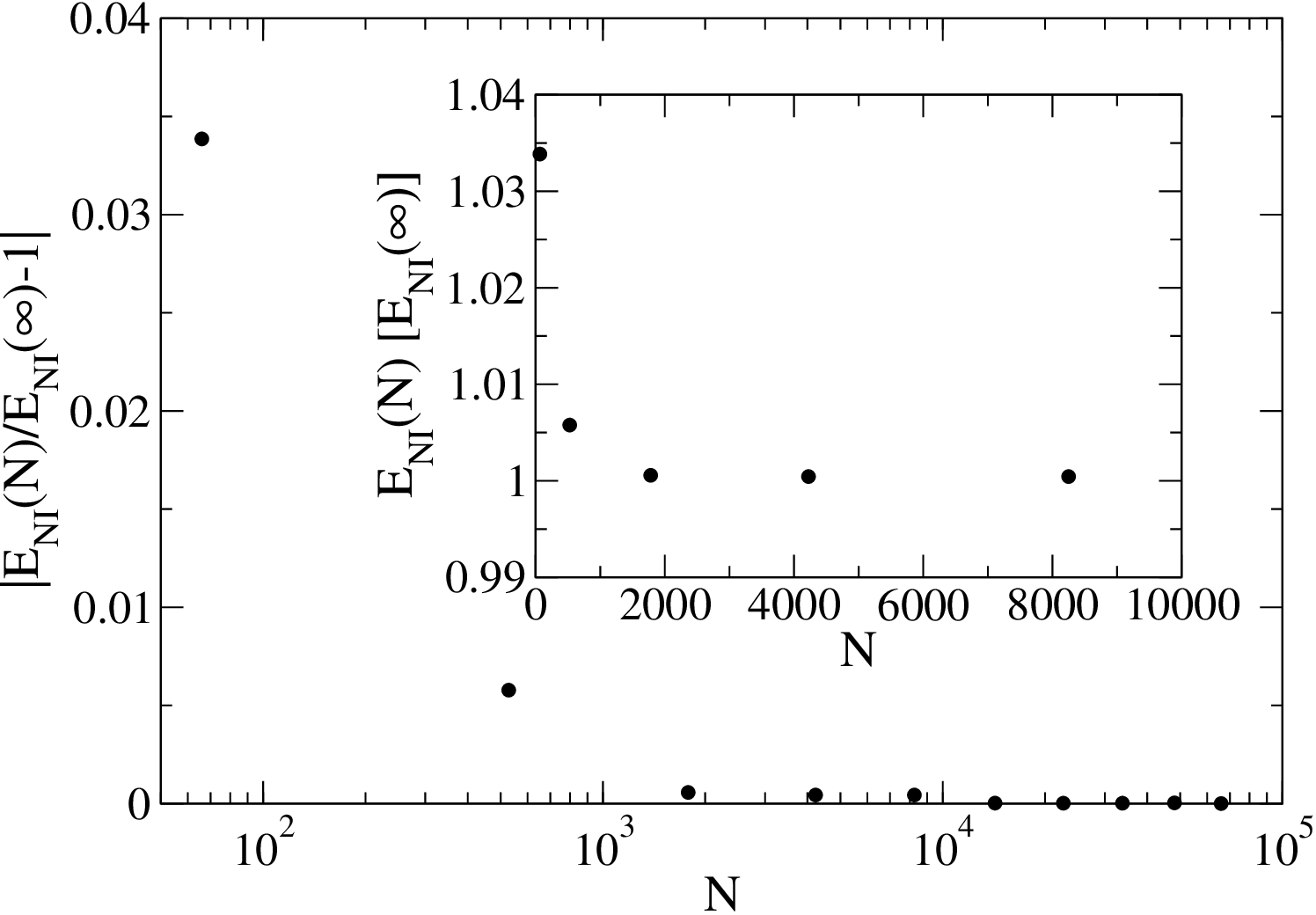}
\caption{FS dependence of the non-interacting Fermi gas in the presence of a one-body potential
of fixed strength $2v_q=0.5E_F$ and periodicity $q=1.4433\,{\rm fm^{-1}}$. $n$ is fixed at  $0.1 \,{\rm fm^{-3}}$.\label{fig:NI_fixed}}
\end{center}
\end{figure}

We now extend the above problem to include our external potential. The Hamiltonian is:
\begin{equation}
\hat{H}=-\frac{\hbar^2}{2m}\sum_{i}{\nabla_i^2}+v_{\rm ext},
\label{eq:NI_H}
\end{equation}
where $v_{\rm ext}=\sum_{i}{v(\mathbf{r}_i)}$ and $v(\mathbf{r}_i)=2v_{q}cos(\mathbf{q}\cdot \mathbf{r}_i)$, as before. 
The orbitals are given by Mathieu functions and the energies by the corresponding characteristic values. Similarly to the free gas, intensive properties converge in the TL. We calculate the energy per particle of the perturbed gas in the same manner as the free gas. Doing so demonstrates convergence in the TL. This is shown in Fig.~\ref{fig:NI} where the line gives energy per particle versus particle number. The density is fixed at $0.1 \,{\rm fm^{-3}}$ and $q$ is set to $4\pi/L$ so that 2 periods of the potential fit inside the box. The amplitude of the potential is $0.5\, E_F = 21.363\, {\rm MeV}$. The curve 
in Fig.~\ref{fig:NI} shows a convergence near 0.8: this is not 1, as could be expected from the presence of the external
potential.

\begin{figure}[t]
\begin{center}
\includegraphics[width=1.0\columnwidth,clip=]{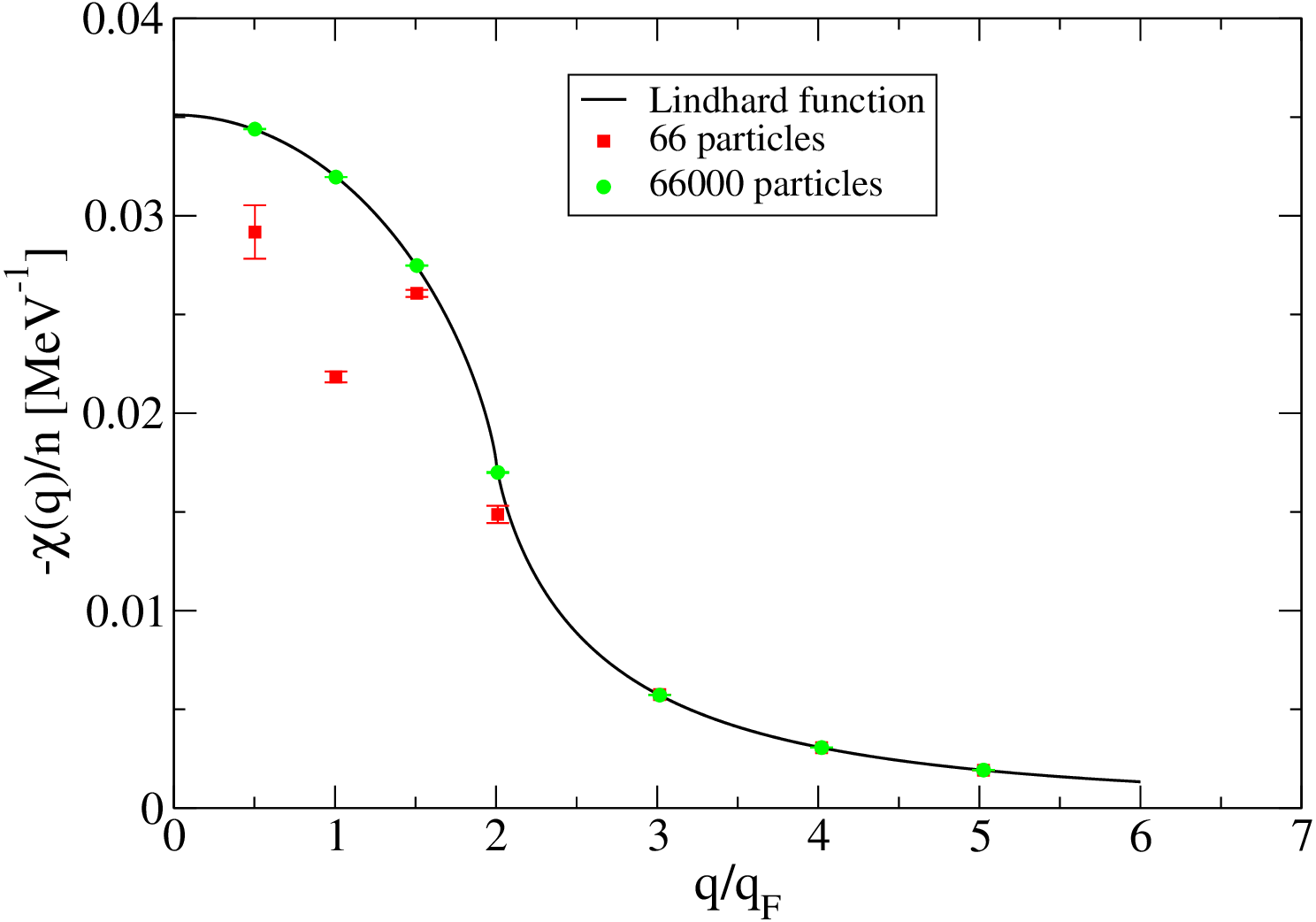}
\caption{Static-response function of the non-interacting free-Fermi gas at a density of $0.1 \,{\rm fm^{-3}}$. The
squares and circles are for 66 and 66000 particles respectively. The line is the Lindhard function describing the
response in the TL.}
\label{fig:NIresponse}
\end{center}
\end{figure}
We use the energy per particle of the non-interacting system to handle FS effects in the interacting problem (see also Ref.~\cite{Kwee:2008}). Our goal is to describe an extended neutron system. To accomplish this, we keep the potential fixed and compute the 
energy per particle versus particle number. Keeping in mind translational invariance, we require an integer number of periods in the box. Thus we choose to perform calculations for a discrete set of particle numbers. This is displayed in Fig.~\ref{fig:NI_fixed} where the density and amplitude are $0.1 \,{\rm fm^{-3}}$ and $21.363\, {\rm MeV}$ respectively. $q$ is fixed so that two periods fit in the box at $N=66$. Of course, the energy per particle converges in the TL. FS effects in the energy per particle $E_I(N)$ of the interacting system will be handled by extrapolating to the TL:
\begin{equation}
E_{\rm I}(\infty)=E_I(N)-E_{NI}(N)+E_{NI}(\infty)
\label{eq:FSfix}
\end{equation}
We tested Eq.~(\ref{eq:FSfix}) by applying it to energy calculations of homogeneous neutron matter. This was done for the SLy4 energies of 66 neutrons at various densities. The results are shown in Table~\ref{table:SLy4}. They agree with SLy4 energies of homogeneous neutron matter to within 0.5\%. This boosts our confidence in Eq.~(\ref{eq:FSfix}).

We further highlight the importance of FS corrections by comparing calculations of the response function to the analytically known response in the TL. The response function for 66000 particles at a density of $0.1 \,{\rm fm^{-3}}$ (circles in Fig.~\ref{fig:NIresponse}) matches the Lindhard function (solid line). This makes sense since 66000 particles is practically in the TL, as per Fig.~\ref{fig:NI_fixed}. The response for 66 particles at this density (squares) does not match the Lindhard function except at large $q$. This stresses that 66 particles is not in the TL so it is important that FS effects be handled in order to study infinite neutron matter.(Note that the FS handling in Ref.~\cite{Buraczynski&Gezerlis:2016} suffered from a numerical error in 
the, near-trivial, calculation for $E_{NI}(66)$; a similar error was present in $E_{NI}(66000)$ but was immaterial there. This error is corrected here.)
We note that the
behavior exhibited by the 66-particle results in Fig.~\ref{fig:NIresponse} is 
observed at other densities also: there is a dip as the $q$ is lowered,
before the response goes back up for the lowest-$q$ point.

\begin{table}[b]
\begin{center}
\caption{SLy4 energies of a free-Fermi gas of neutrons computed using the local density approximation. Eq.~(\ref{eq:FSfix}) was used to extrapolate to the TL and compare to SLy4 energies of homogeneous neutron matter.\label{table:SLy4}}
\begin{tabular}{lccc}
\hline
$n\,(\rm fm^{-3})\,\,\,\,\,\,\,$ & $E_I(66)\,\rm{(MeV)}$\,\,\,\,\,\,\, & $E_I(\infty)\,\rm{(MeV)}$ \,\,\,\,\,& TL\,\rm{(MeV)} \\
\hline
$0.04$ & \,\,\,$7.15$ &\,\, $7.22$ & \,\,\,$7.23$ \\
$0.06$ & \,\,\,$8.62$ & \,\,\,\,\,$8.71$ & \,\,\,$8.73$ \\
$0.08$ &\,\,\,$9.99$ & \,\,\,\,\,\,\,$10.10$ & \,\,\,\,\,$10.13$ \\
$0.10$ & $\,\,\,\,\,11.40$& \,\,\,\,\,\,\,$11.53$ & \,\,\,\,\,$11.58$ \\
\hline
\end{tabular}
\end{center}
\end{table}

\section{Interacting problem}
\label{sec:results}

\subsection{Equation of state}

We computed the equation of state of 66 neutrons both with and without an external potential. We first present results that do not include NNN interactions. The NN interactions are given by the Argonne v8' potential \cite{Wiringa:2002}. Calculations were performed for densities in the range of $0.02 \,{\rm fm^{-3}}$ to $0.12\, {\rm fm^{-3}}$ since these are similar to the densities found in the crust and outer-core of a neutron star.
 \begin{figure}[t]
\begin{center}
\includegraphics[width=1.0\columnwidth,clip=]{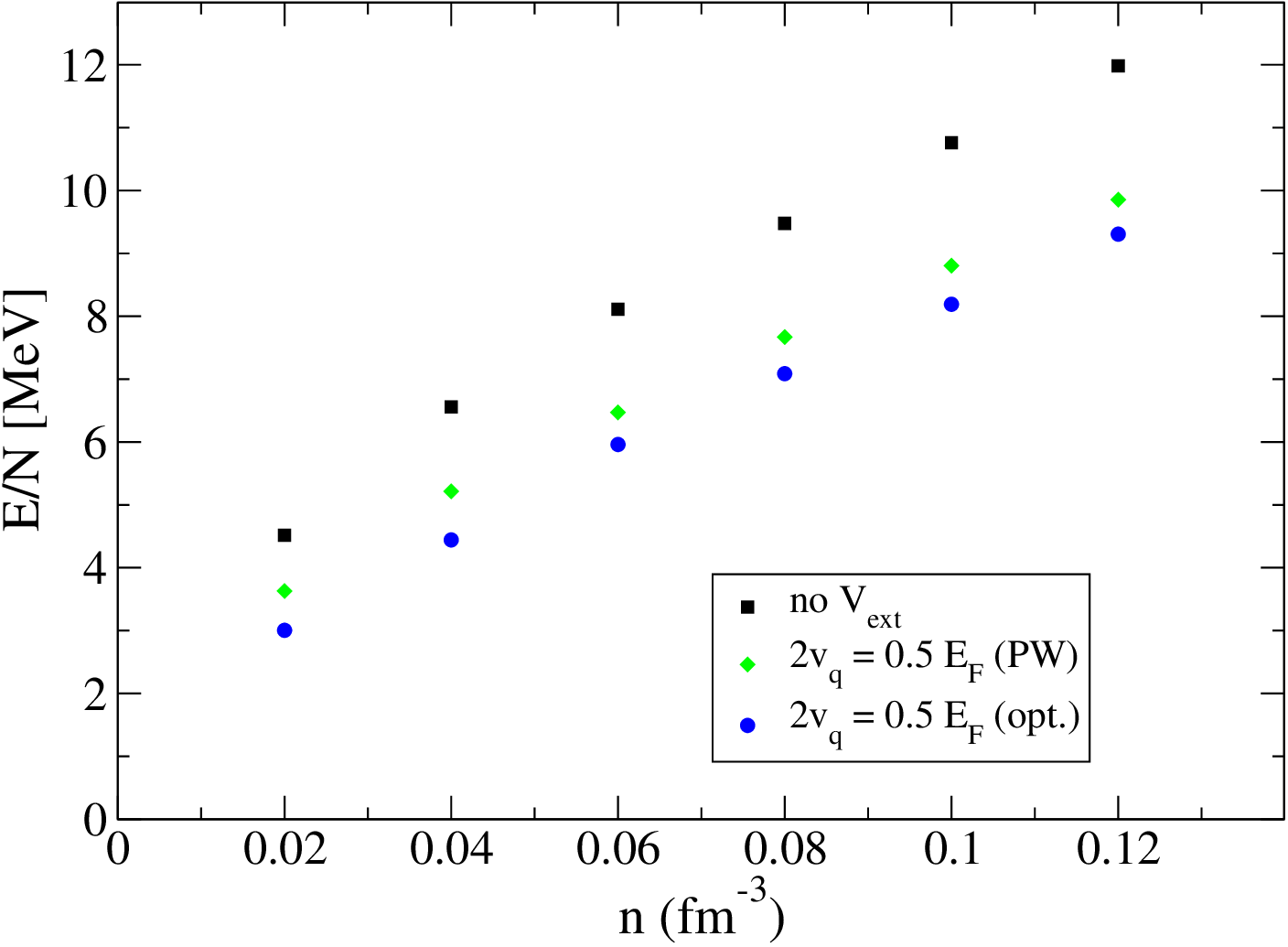}
\caption{AFDMC neutron-matter energy per particle for 66 particles 
as a function of density using only NN interactions. Squares denote the case without a one-body potential; diamonds a one-body potential of fixed strength $2v_q=0.5E_F$, periodicity $q=4\pi/L$, and plane-wave single-particle orbitals; circles a one-body potential of fixed strength $2v_q=0.5E_F$, periodicity $q=4\pi/L$, and optimized single-particle orbitals (opt.).\label{fig:Evsn}}
\end{center}
\end{figure}
The nodal structure of the trial wave function is important, so the single-particle orbitals used in the Slater determinant must be chosen carefully. We use the solutions of the one-body problem with the same external potential and no interactions. For the case of no external potential these are the plane-waves of the non-interacting free-Fermi gas. The AFDMC results for these are the squares in Fig.~\ref{fig:Evsn}. The energy increases with increasing density. The energies agree with known values~\cite{Maris:2013}. We computed energies using plane-wave orbitals for a one body potential of strength $2v_q=0.5 E_F$ and two periods of the potential in the box (diamonds). This yields lower energies than the unperturbed problem. However, this is not good enough since we have not optimized the trial wave-function. The energies for optimized single-particle orbitals (circles) are up to ~1 MeV different from the plane-wave results. Overall, the AFDMC energies with an external potential are several MeV smaller than those without an external potential.  
\begin{figure}[t]
\begin{center}
\includegraphics[width=1.0\columnwidth,clip=]{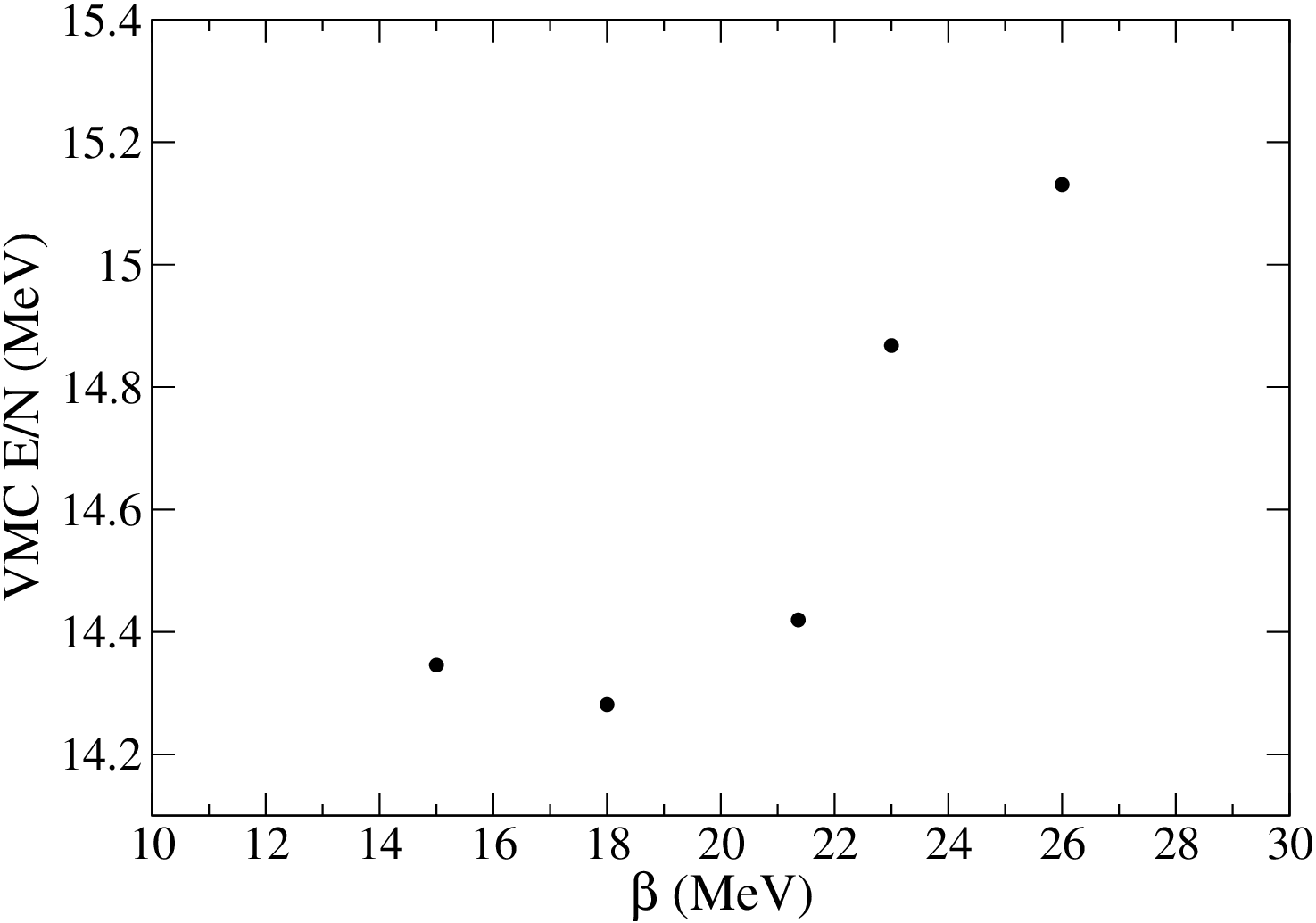}
\caption{Neutron-matter VMC energy per particle as a function of the variational parameter $\beta$ using NN interactions.$\,n_0=0.1 \,{\rm fm^{-3}}$, $2v_q=0.5E_F=21.363$ MeV, and $q=4\pi/L$. \label{fig:beta}}
\end{center}
\end{figure}
This reflects the particles' tendency to stay away from the repulsive regions of the potential and collect in the wells of the potential. Note that the amplitude of the external potential used is density dependent. Also, the period of the potential decreases with increasing density. This is consistent with what happens in a neutron-star crust where the lattice spacing decreases with increasing density.

We now describe the optimization procedure. Firstly, Mathieu functions yield lower VMC energies than plane-waves in the Slater determinant. Since this is a variational optimization the goal is to minimize the VMC energy. We do this using a variational parameter $\beta$ where the solutions to the one-body non-interacting problem with external potential $v(\mathbf{r})=\beta \cos(\mathbf{q} \cdot \mathbf{r})$ \cite{Moroni:1992} are used in the Slater determinant. For the case $\beta=2v_q$ these are the orbitals of the one-body non-interacting problem with the same external potential as the system under study. Consider a density of $0.1 \,{\rm fm^{-3}}$ for the case where $2v_q=0.5 E_F$ and two periods of the potential fit in the box. The minimum VMC energy that we simulated occurs at $\beta = 18$ MeV (Fig.~\ref{fig:beta}). The AFDMC energy for $\beta=18$ MeV is 8.19 MeV. The AFDMC energy for $\beta=2v_q$ is 8.15 MeV. Thus, the difference in energy due to the $\beta$ optimization procedure is much smaller than the difference due to using plane-waves vs Mathieu functions. Therefore, it is safe to 
set $\beta=2v_q$ for practical purposes.
\begin{figure}[t]
\begin{center}
\includegraphics[width=1.0\columnwidth,clip=]{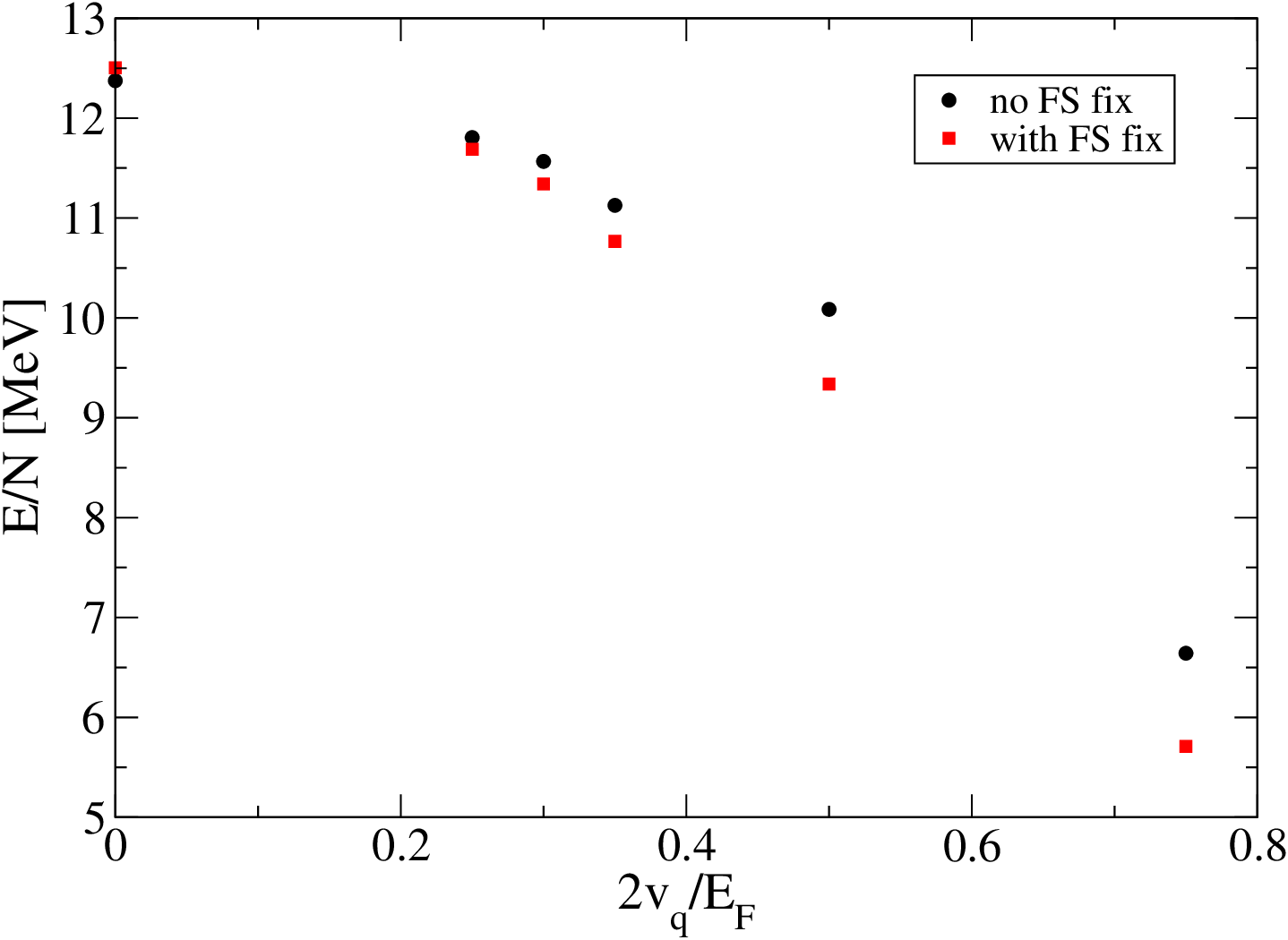}
\caption{Neutron-matter energy per particle as a function of the one-body potential strength using NN+NNN interactions and AFDMC. $\,n_0=0.1 \,{\rm fm^{-3}}$ and $q=4\pi/L$. Circles correspond to energies prior to handling FS effects and squares to energies extrapolated to the TL.\label{fig:Evsvq}}
\end{center}
\end{figure}
\begin{figure}[b]
\begin{center}
\includegraphics[width=\columnwidth]{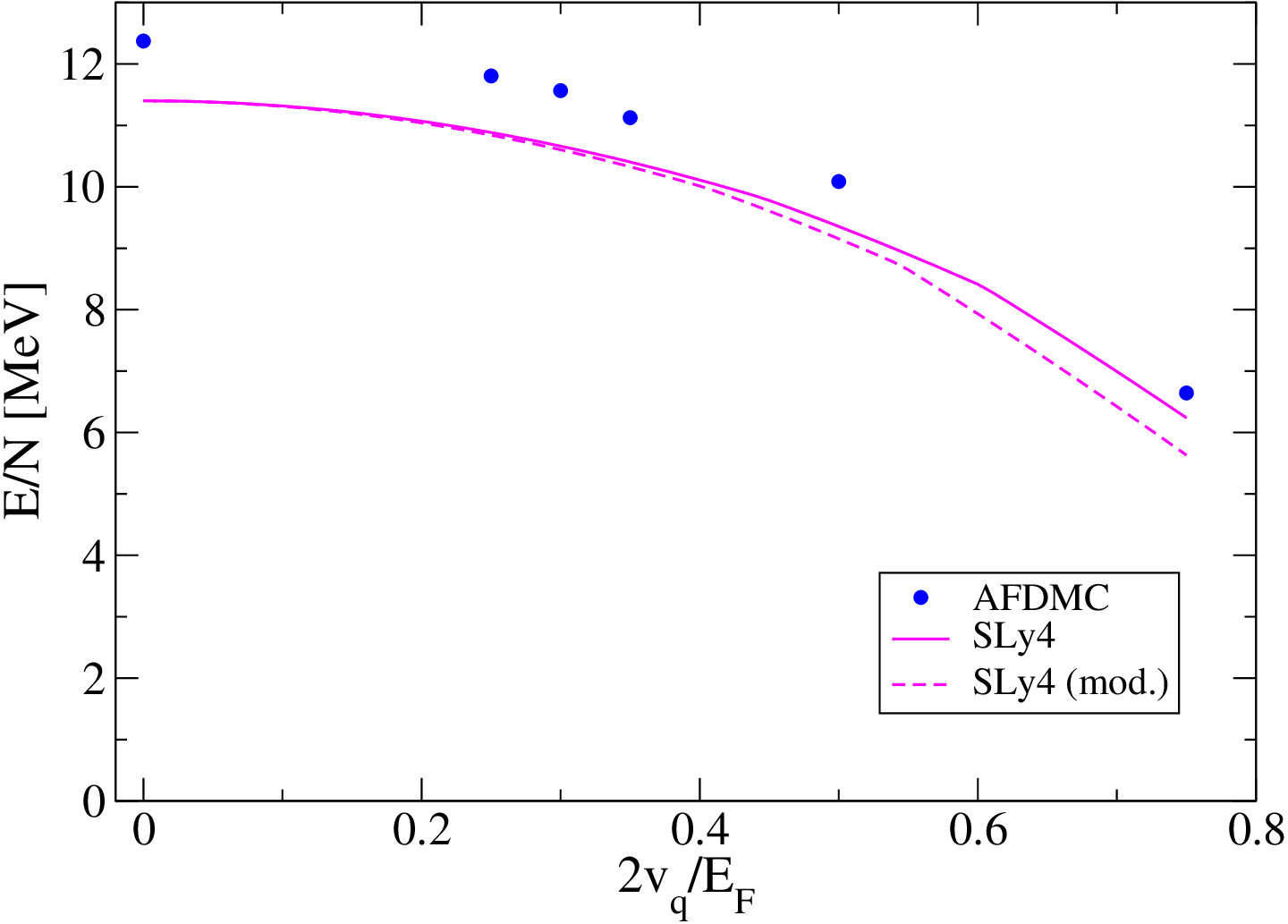}
\caption{Neutron-matter energy per particle for 66 particles as a function of the one-body potential strength
at a density of $n_0=0.10\, {\rm fm^{-3}}$, using NN+NNN interactions and a 
one-body potential periodicity $q=4\pi/L$. Circles denote AFDMC results,
and the solid line follows from the SLy4 energy-density functional. 
The dashed line corresponds to SLy4 results with a modified isovector gradient term. 
\label{fig:ener_v_stren}}
\end{center}
\end{figure}

We are primarily interested in the static response of neutron matter. This requires us to perform calculations across a set of strengths of the external potential. It is important that the strengths are large enough so that the energy is statistically different from homogeneous neutron matter. At the same time the filling of the single particle orbitals changes at larger $v_q$ in comparison
to the free non-interacting orbital filling. We performed simulations for various orbital fillings and found no significant change in the energy. We did this for $2v_q=0.5 E_F$ at $n=0.08\,{\rm fm^{-3}}$ for two different fillings and found AFDMC energies of $7.072$ and $7.062$ MeV. The same was done for  $2v_q=0.75 E_F$ at $n=0.04\,{\rm fm^{-3}}$ yielding $1.818$  and $1.822$ MeV. We decided to calculate energies for $2v_q=0.25,\,0.3,\,0.35,\,0.5,\,0.75 \,E_F$. NNN interactions were included in these simulations. Specifically, the Urbana IX potential was used, as is appropriate for neutron-rich matter \cite{Pudliner:1997}. Turning up the potential strength results in a decrease in energy. This is seen in the circles in Fig.~\ref{fig:Evsvq} for 66 particles at a density of $0.1 \,{\rm fm^{-3}}$ and two periods of the potential in the box. We have also applied Eq.~(\ref{eq:FSfix}) to extrapolate to the TL (squares). We extract the response function from such results. We present those results in the next section.

\subsection{Constraining the isovector gradient coefficient}

We have also used the response of neutron matter to constrain energy density functionals. The energy in Eq.~(\ref{eq:Eint}) is minimized with respect to the variational parameter $\beta$ and different orbital fillings. This is done for a range of potential strengths.
\begin{figure}[t]
\begin{center}
\includegraphics[width=\columnwidth]{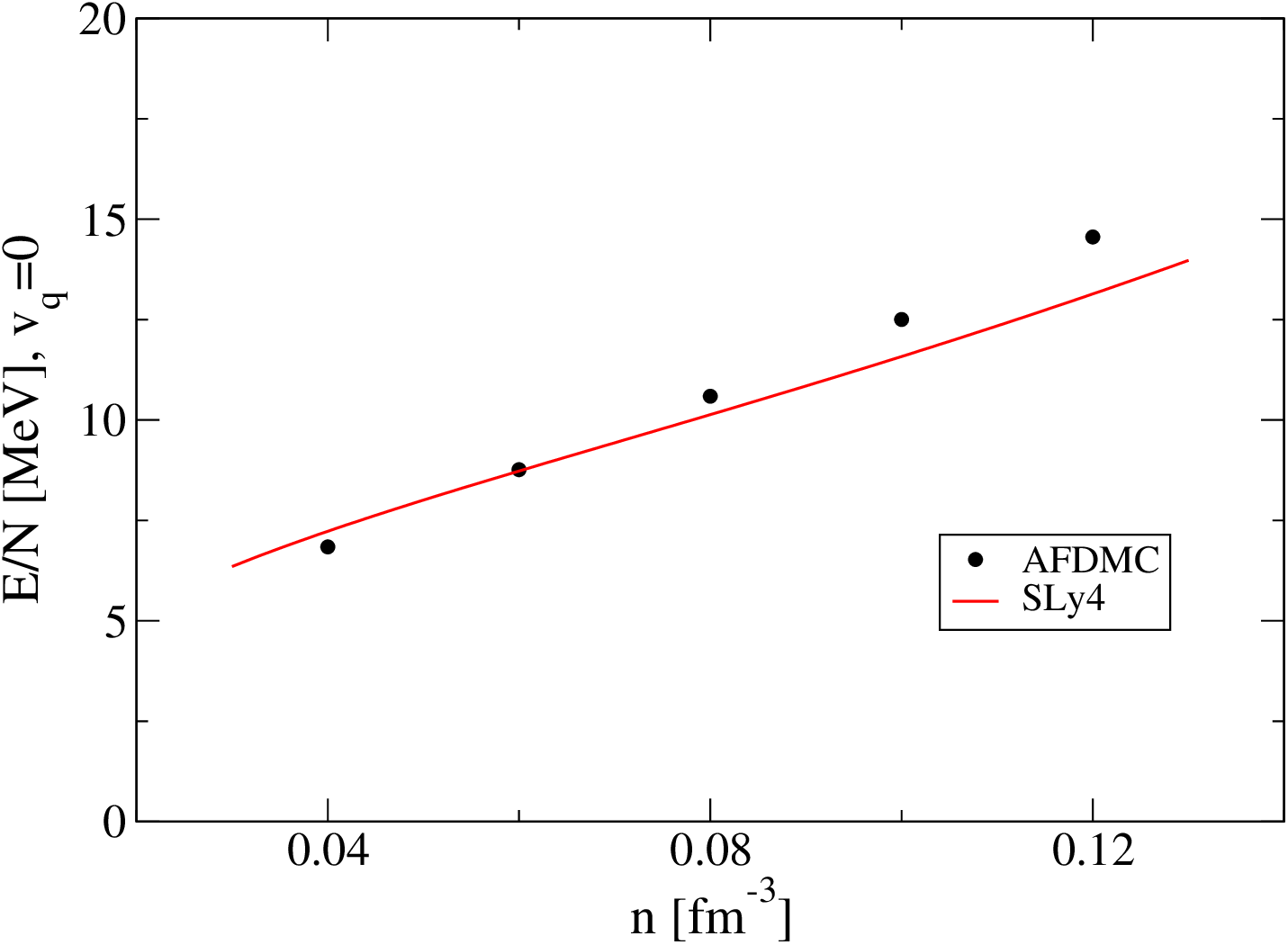}
\caption{TL AFDMC and TL SLy4 results in the absence of a one-body potential. 
\label{fig:homogeneous}}
\end{center}
\end{figure}
Fig.~\ref{fig:ener_v_stren} displays the results of this procedure for SLy4 (solid line) for a 66 particle system at a density of $0.10\, {\rm fm^{-3}}$ and two periods of the potential in the box. The 66 particle AFDMC NN+NNN energies (circles) are more repulsive than the SLy4 energies. The separation between the AFDMC and SLy4 energies is largest at $v_q=0$. The separation decreases as $v_q$ increases. $\nabla n=0$ at homogeneity so the isovector gradient term does not contribute to the $v_q=0$ energy. Thus the difference between the AFDMC and SLy4 energies at homogeneity is due to the bulk energy of the system. This homogeneous mismatch in energy must be respected in fitting the EDF to AFDMC results. The isovector gradient term has the effect of bringing the SLy4 energies closer to the AFDMC energies at larger $v_q$. 
\begin{figure}[t]
\begin{center}
\includegraphics[width=\columnwidth]{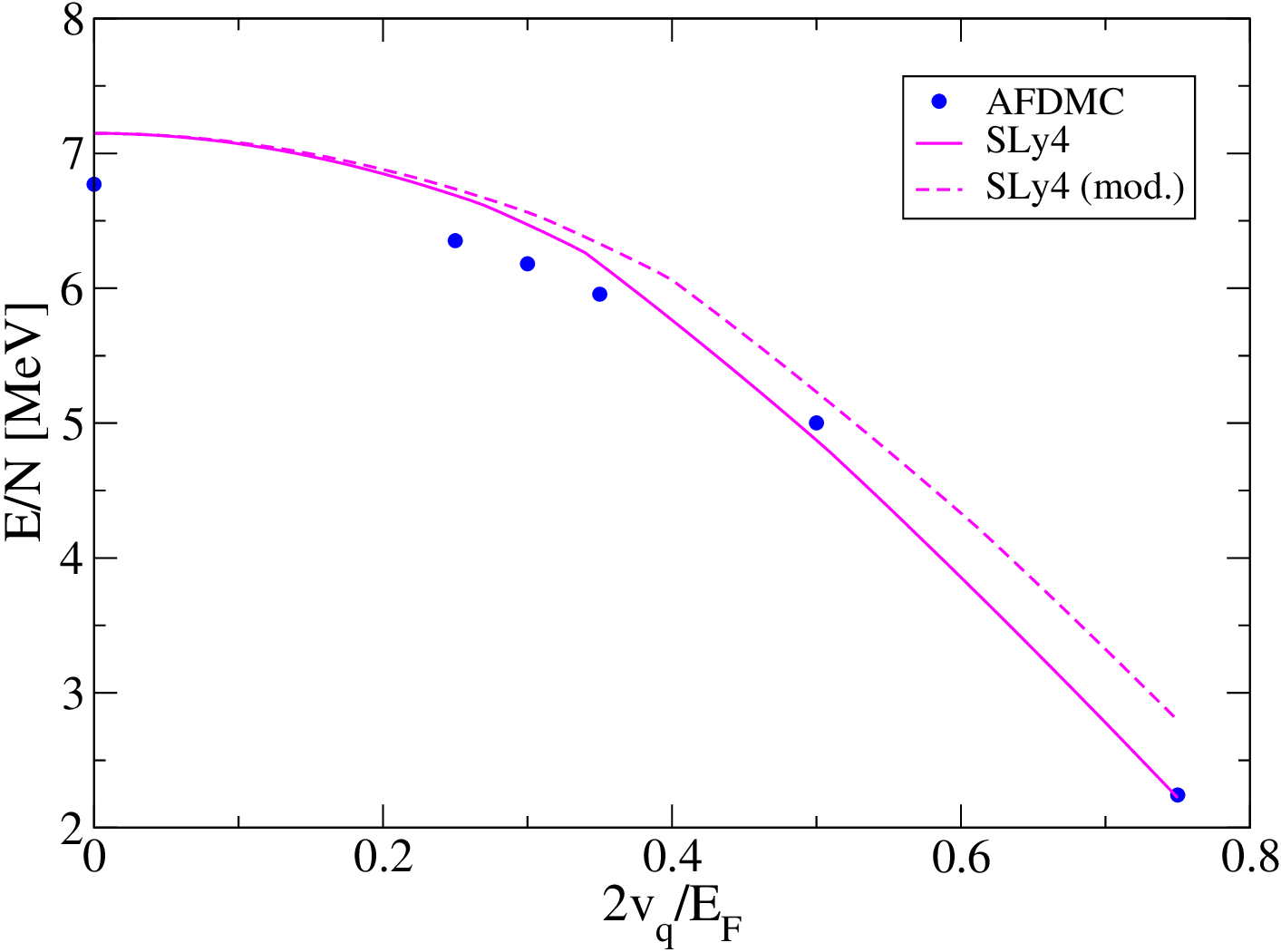}
\caption{Neutron-matter energy per particle for 66 particles as a function of the
one-body potential strength
at a density of $n_0=0.04\, {\rm fm^{-3}}$, using NN+NNN interactions and a 
one-body potential periodicity $q=4\pi/L$. Circles denote AFDMC results,
and the solid line follows from the SLy4 energy-density functional. 
The dashed line corresponds to SLy4 results with a modified isovector gradient term.
\label{fig:ener_v_stren0.04}}
\end{center}
\end{figure}
Thus, our fitting of the isovector gradient term maintains the $v_q=0$ difference between the EDF and QMC energies. We fit to low strengths $2v_q=0.25 \,\,\rm{and}\,\, 0.3 E_F$, 
in order to ensure that the density perturbation magnitude 
is not comparable to the unperturbed density. We found a modified isovector gradient term of $C^{\Delta n}_1=-29 \,\rm{MeV \,fm^5}$ at this density (dashed line in Fig.~\ref{fig:ener_v_stren}).

The homogeneous energy difference between AFDMC and SLy4 impacts how the isovector gradient term should be modified. Above $n=0.06\,\rm{fm^{-3}}$ the homogeneous AFDMC energy (circles in Fig.~\ref{fig:homogeneous}) is more repulsive than the homogeneous SLy4 energy (solid line in Fig.~\ref{fig:homogeneous}). Below this density the AFDMC is more attractive than the SLy4 energy. We see this at $n=0.04\,\rm{fm^{-3}}$ where the AFDMC NN+NNN energies (circles in Fig.~\ref{fig:ener_v_stren0.04}) are smaller than SLy4 (solid line in Fig.~\ref{fig:ener_v_stren0.04}) for small enough $v_q$. The separation between AFDMC and SLy4 decreases with increasing $v_q$ at both densities larger and smaller than $0.06\,\rm{fm^{-3}}$. This means that the fitted isovector gradient term is density dependent. This result is also found using the density-matrix expansion~\cite{Holt:2011,Kaiser:2012b}. For $n>0.06\,\rm{fm^{-3}}$ the isovector gradient fit must make the SLy4 energies more attractive if they are to be equidistant from the AFDMC energies. This requires a decrease in the isovector gradient term. For $n<0.06\,\rm{fm^{-3}}$ an increase in the isovector gradient term is required. At a density of  $0.04\,\rm{fm^{-3}}$ we find a modified isovector term of $C^{\Delta n}_1=9 \,\rm{MeV \,fm^5}$ (dashed line in Fig.~\ref{fig:ener_v_stren0.04}). Since we fit the isovector term to low strengths, the modified SLy4 still approaches the AFDMC results at some large $v_q$.

\begin{figure}[t]
\begin{center}
\includegraphics[width=\columnwidth]{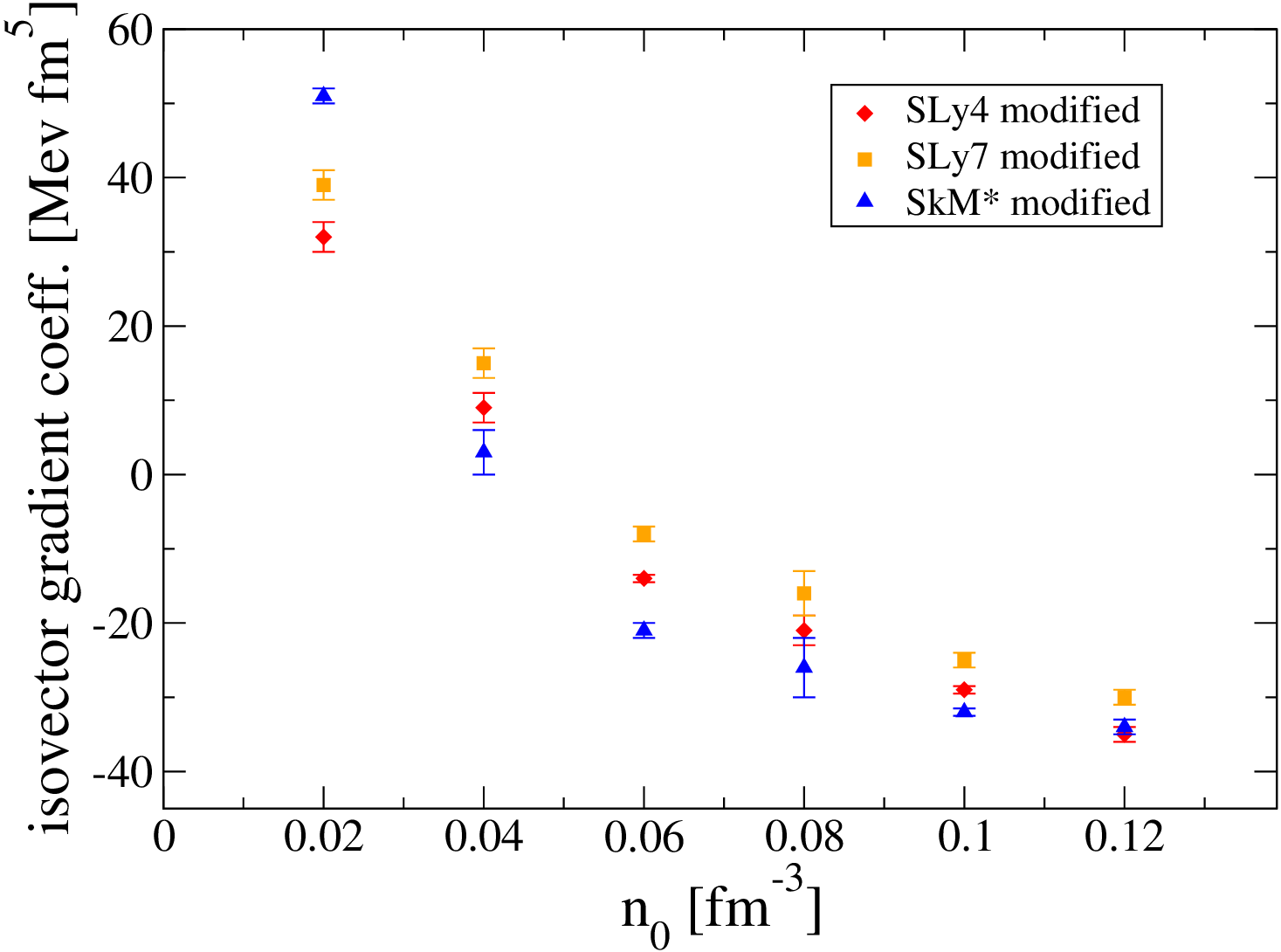}
\caption{Isovector gradient coefficients for the modified SLy4, SLy7, and SkM* Skyrme potentials. All calculations were done with two periods of the potential in the box. Coefficients were modified to make Skyrme responses match
the QMC responses. \label{fig:3funcs}}
\end{center}
\end{figure}

We have carried out calculations such as those in Fig.~\ref{fig:ener_v_stren} and 
Fig.~\ref{fig:ener_v_stren0.04} for several other densities. They exhibit qualitatively
the same trends as discussed above. We then used these AFDMC results
to constrain the isovector coefficient for several functionals. Specifically, 
Fig.~\ref{fig:3funcs} lists the isovector gradient term of modified SLy4, SLy7, and SkM*. All of these were done using two periods of the potential in the box. The errors were determined by fitting to different strengths and examining the spread of the modified isovector gradient term. Note that the density dependent energy versus $v_q$ behaviour described above does not hold for SkM*. Nevertheless, our fitting prescription yields the same density dependence in the modification of the isovector term for all three parameterizations. We see a decrease
is required at large densities. At low densities there is an increase in the isovector term. The isovector term is least modified at $n=0.06\,\rm{fm^{-3}}$, where in 
the homogeneous case the AFDMC and SLy4 results agree reasonably well, as per
Fig.~\ref{fig:homogeneous}.

Note that the isovector coefficient fits discussed above were all carried out 
by focusing on the EDF and QMC results for $L=2d$, i.e., 
using two periods of the potential in the box. As we will see in the following subsection,
our attempt to make the QMC and EDF results equidistant essentially amounts to trying to
match (not QMC energies to EDF energies, but) the EDF response function to the QMC
response function. In other words, we are attempting to modify the SLy4 response 
from Fig.~\ref{fig:SLy40.1resp} below to match that in Fig.~\ref{fig:AFDMC0.1resp} below.
All this, for the specific case of $L=2d$. As the difference between SLy4 and modified SLy4 in Fig.~\ref{fig:SLy40.1resp} shows, for $L=2d$ we would need a more attractive 
modification, whereas for $L=d$ we would need a more repulsive one (and for $L=3d$
we would need a modification that is more attractive than that for $L=2d$). 
Thus, the optimal
thing to do is to try to optimize results for as many periodicities as possible
simultaneously. We have done this at the two densities of  $n=0.10\,\rm{fm^{-3}}$ and $n=0.04\,\rm{fm^{-3}}$, where we have produced AFDMC results for many different 
periodicities, as discussed below.

\subsection{Response functions}

\begin{figure}[t]
\begin{center}
\includegraphics[width=\columnwidth]{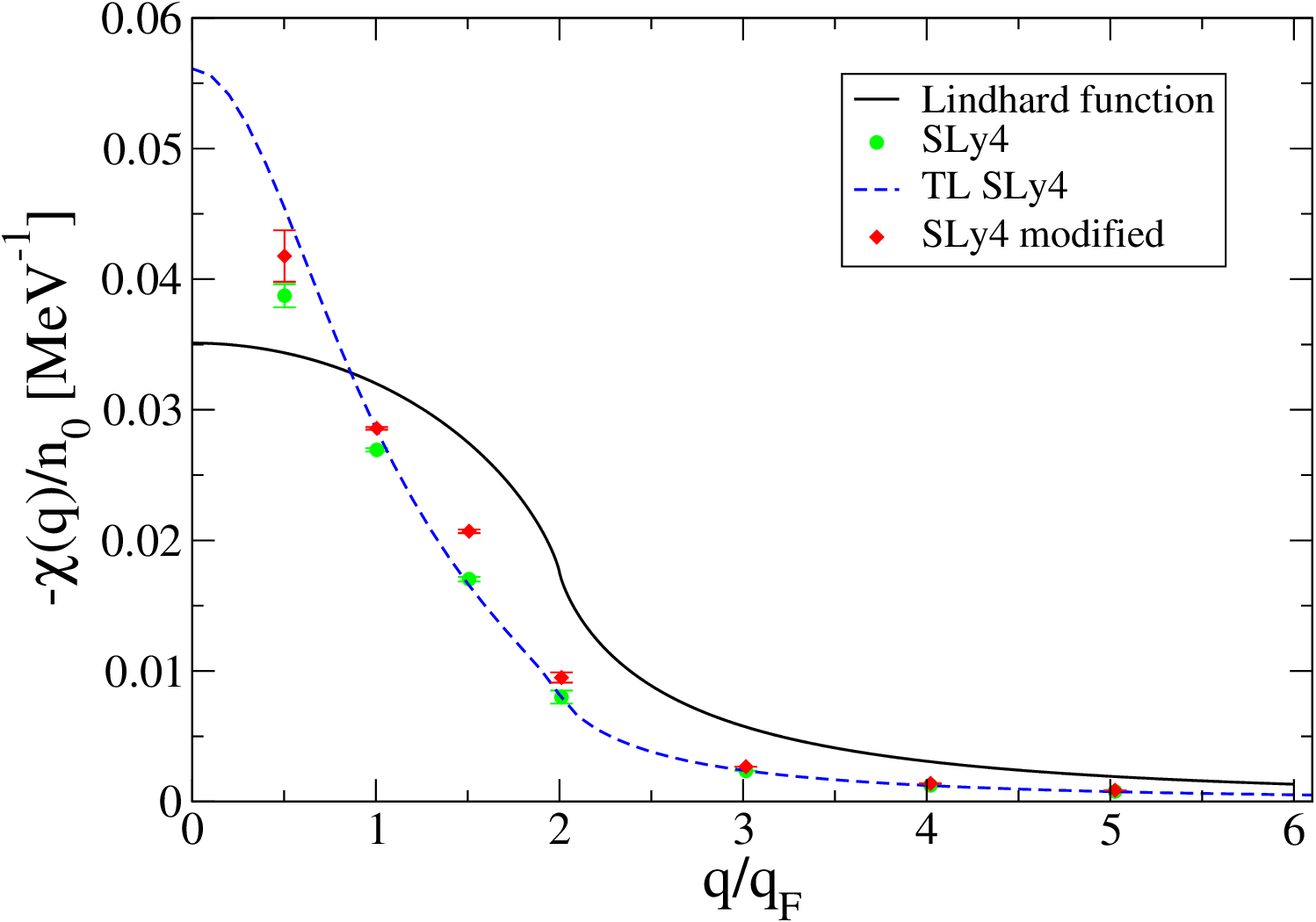}
\caption{Static-response functions of neutron-matter at a density of $0.10\,\rm{fm^{-3}}$. The circles are the SLy4 response extrapolated to the TL limit. The diamonds are for the modified SLy4 with $C^{\Delta n}_1=-29 \,\rm{MeV \,fm^5}$ extrapolated to the TL. The response was extracted by fitting to $2v_q=0.25,\,0.3,\,0.35,\,0.5,\,$ and $0.75\, E_F$. The dashed curve is the SLy4 response produced in the TL~\cite{LacroixPrivate}. The solid line is the Lindhard function describing the response of a non-interacting Fermi gas.\label{fig:SLy40.1resp}}
\end{center}
\end{figure}

Using calculations like those discussed above, we extracted linear density-density response functions at both $0.04\,\rm{fm^{-3}}$ and $0.10\,\rm{fm^{-3}}$ for AFDMC, SLy4 and modified SLy4 results. (To do this, we fit to even-powered polynomials up to fourth order in Eq.~(\ref{eq:energy}).) Since we are studying neutron matter we use Eq.~(\ref{eq:FSfix}) to extrapolate energies to the TL. It was previously mentioned that we only consider $q$ such that an integer number of periods of the potential fit in the box. We have performed simulations for $q=$ 2, 4, 6, 8, 12, 16, 20 times $\pi/L$ corresponding to 1, 2, 3, 4, 6, 8, and 10 periods of the potential inside the box. 

\begin{figure}[b]
\begin{center}
\includegraphics[width=\columnwidth]{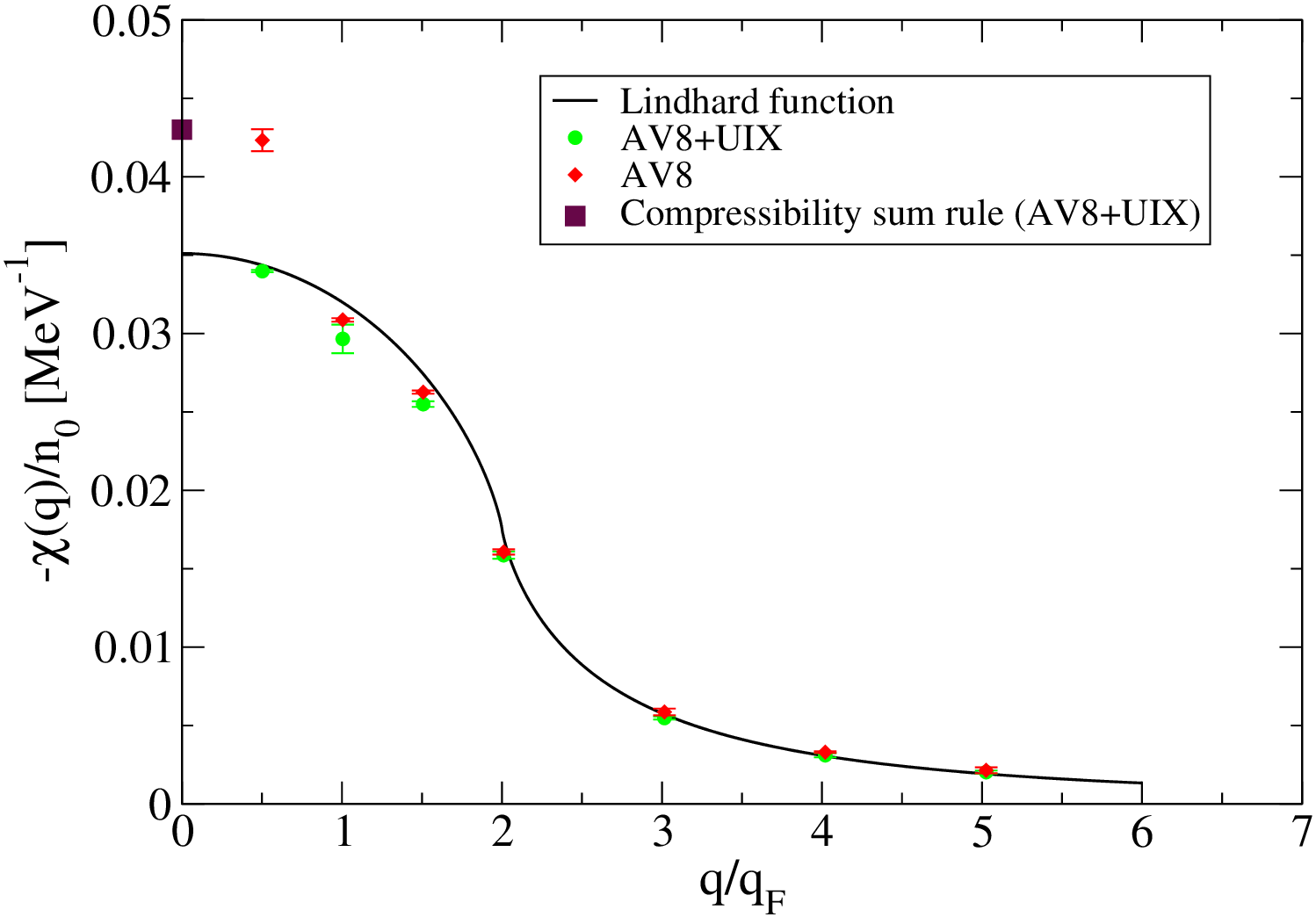}
\caption{Static-response functions of neutron-matter at a density of $0.10\,\rm{fm^{-3}}$. Produced using AFDMC results. Circles are with NN+NNN interactions extrapolated to the TL limit. Diamonds are for NN interactions extrapolated to the TL. The response was extracted by fitting to: $2v_q=0.25,\,0.3,\,0.35,\,0.5,\,$ and $0.75\, E_F$. The square is the response at $q=0$ predicted by the compressibility sum rule for the NN+NNN case. The curve is the Lindhard function describing the response of a non-interacting Fermi gas.\label{fig:AFDMC0.1resp}}
\end{center}
\end{figure}

We have also taken advantage of the compressibility sum rule: this provides us 
with a way to calculate $\chi (0)$ starting from the energy per particle as a function of density of the unperturbed system:
\begin{equation}
\frac{1}{\chi(0)}=-\frac{\partial^2(n_0E/N)}{\partial n_0^2}
\label{eq:CSR}
\end{equation}
We used Eq.~(\ref{eq:CSR}) to compute $\chi(0)$ for the various response functions that we extracted and checked for consistency with our modulated results.

We first present results at  $n_0=0.10\,\rm{fm^{-3}}$.
We have found that the SLy4 response (circles in Fig.~\ref{fig:SLy40.1resp}) does not match the Lindhard function (solid line in Fig.~\ref{fig:SLy40.1resp}) although there are similarities. Both responses have a finite $\chi(0)$ and go to 0 at large $q$. We compare to the SLy4 response function in the TL~\cite{LacroixPrivate} (dashed line in Fig.~\ref{fig:SLy40.1resp}). Our response agrees with it pretty well for all except the smallest $q$ values. The compressibility sum rule gives a value of $-\chi(0)/n_0=0.057\, \rm{MeV^{-1}}$ for SLy4 which matches the TL SLy4.
We are also interested in the modified SLy4 response (diamonds in Fig.~\ref{fig:SLy40.1resp}). This response is similar in shape but larger in magnitude than the SLy4 response we extracted. This makes sense since the modified SLy4 is more attractive than SLy4. The compressibility sum rule gives the same $\chi(0)$ for modified SLy4 as SLy4 since the unperturbed system is independent of the gradient term. 

In Fig.~\ref{fig:AFDMC0.1resp} we show updated AFDMC NN+NNN results (circles):
note that these include the corrected FS handling 
and therefore differ (in the lowest-$q$ response value) 
from the circles in Fig. 3 of Ref.~\cite{Buraczynski&Gezerlis:2016}. The diamonds in Fig.~\ref{fig:AFDMC0.1resp} show the AFDMC NN response function in the TL. Our results in the TL are similar in shape to the Lindhard function (line in Fig.~\ref{fig:AFDMC0.1resp}). At larger $q$ the response goes to zero and matches the Lindhard function. At smaller $q$ the NN+NNN response is smaller than the Lindhard function. The NN response is larger than the Lindhard function at the lowest $q$ value, but other than that it is very similar to the NN+NNN response. The compressibility sum rule gives a value of $-\chi(0)/n_0=0.043\, \rm{MeV^{-1}}$ for neutron matter with our NN+NNN interactions (square in Fig.~\ref{fig:AFDMC0.1resp}) and a value of $0.089\, \rm{MeV^{-1}}$ for NN interactions only. These are larger than the corresponding value of $0.035\, \rm{MeV^{-1}}$ for the Lindhard function. Note that Fermi liquid theory yields $-\chi(q)/n_0 \approx0.035 \,\rm{MeV^{-1}}$ at $n_0=0.10\,\rm{fm^{-3}}$ \cite{Iwamoto:1982,LandauParameters}. 
\begin{figure}[t]
\begin{center}
\includegraphics[width=\columnwidth]{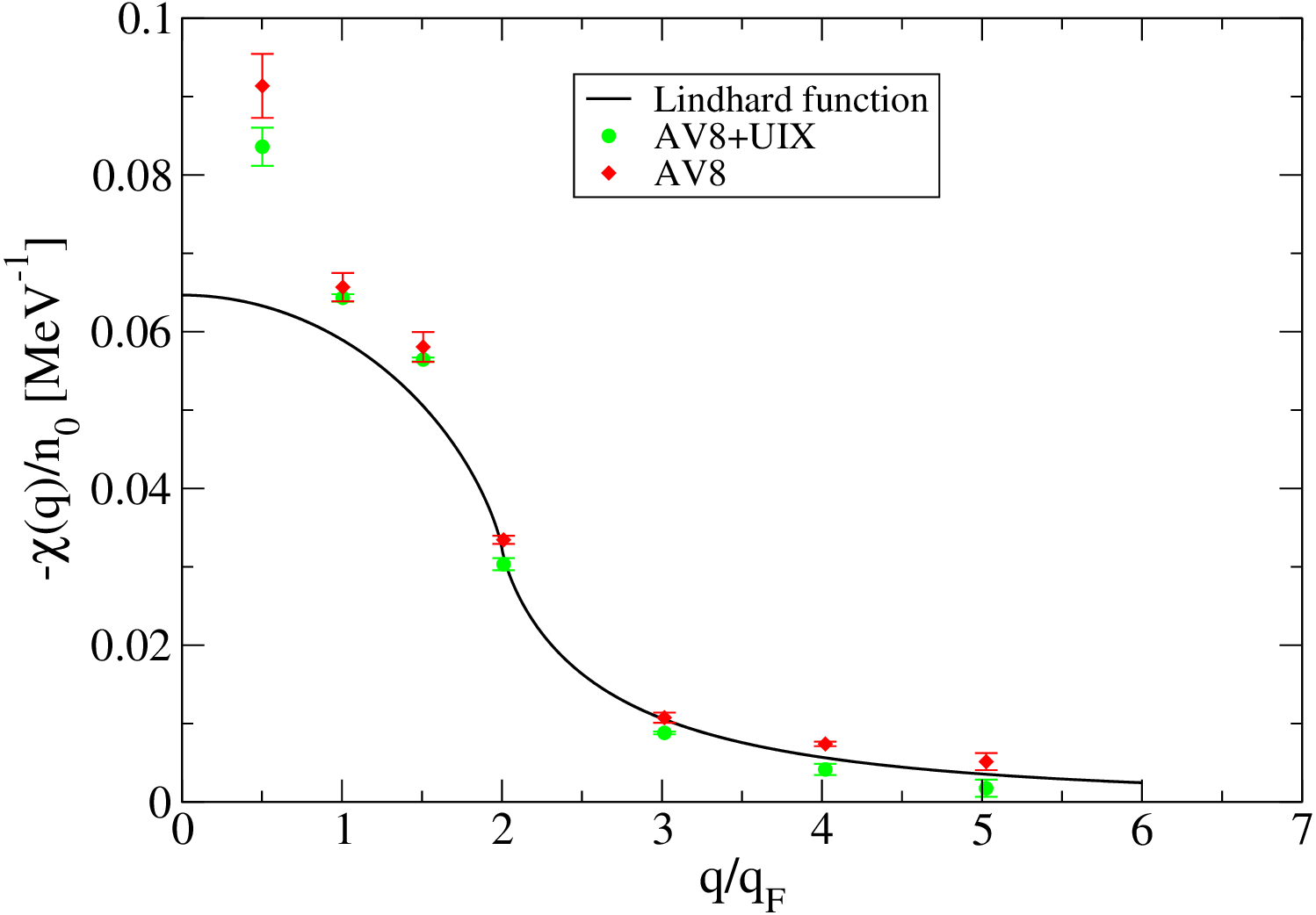}
\caption{Static-response function of neutron-matter at a density of $0.04\,\rm{fm^{-3}}$. The circles are with NN+NNN interactions extrapolated to the TL. Diamonds are for NN interactions extrapolated to the TL. The AFDMC responses were extracted by fitting to: $2v_q=0.25,\,0.3,\,0.35,$ and $0.5\, E_F$. The solid line is the Lindhard function describing the response of a non-interacting Fermi gas. \label{fig:AFDMC0.04resp}}
\end{center}
\end{figure}

We now examine some of the responses at a density of $0.04\,\rm{fm^{-3}}$. We have found that the AFDMC results do not follow the Lindhard function (solid line in Fig.~\ref{fig:AFDMC0.04resp}) as well as the $0.10\,\rm{fm^{-3}}$ AFDMC results do. Both the AFDMC NN response (diamonds in Fig.~\ref{fig:AFDMC0.04resp}) and the AFDMC NN+NNN response (circles in Fig.~\ref{fig:AFDMC0.04resp}) are larger than the Lindhard function at small $q$. The compressibility sum rule gives a value of $-\chi(0)/n_0=0.14\, \rm{MeV^{-1}}$ for neutron matter with our NN+NNN interactions and  $0.19\, \rm{MeV^{-1}}$ for NN interactions only. These are larger than the $-\chi(0)/n_0=0.065\, \rm{MeV^{-1}}$ of the Lindhard function. Fermi liquid theory yields $-\chi(q)/n_0 \approx 0.083 \,\rm{MeV^{-1}}$ 
at $n_0=0.04\,\rm{fm^{-3}}$ \cite{Iwamoto:1982,LandauParameters}.

For all results it was found that the response function goes to zero as $q$ goes to infinity and $\chi(0)$ is finite. In addition, the response functions extracted from AFDMC and the Lindhard functions show more similarity to one another than to the SLy4 responses. Overall, the SLy4 responses are narrower and steeper than the other responses. It is also interesting to contrast these to the response functions of $\rm {}^4He$ \cite{Moroni:1992} and the 3D electron gas \cite{Moroni:1995} both of which have $\chi(0)=0$.  

\begin{figure}[t]
\begin{center}
\includegraphics[width=\columnwidth]{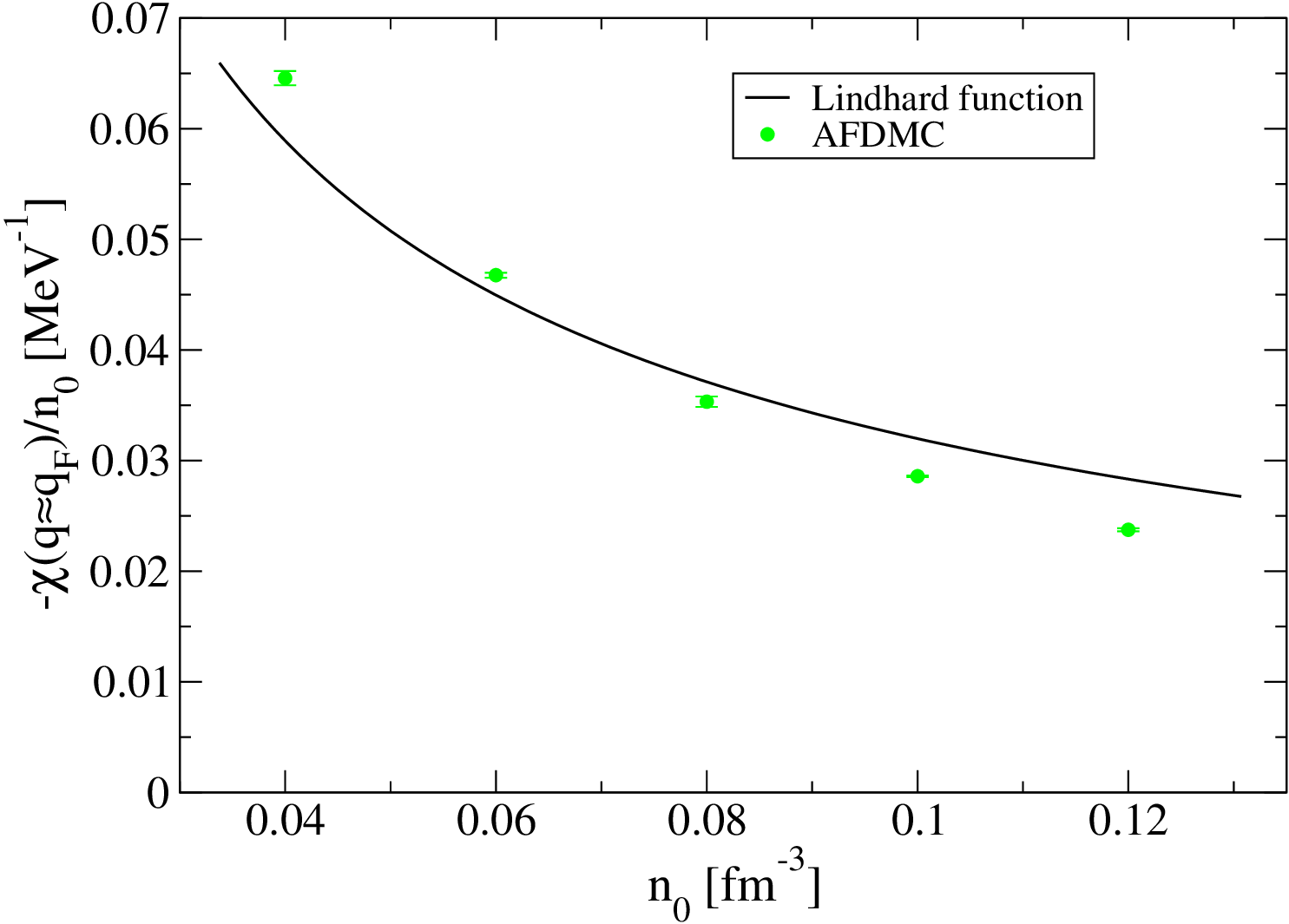}
\caption{Static-response function of neutron-matter across several densities. The circles correspond to 
AFDMC results with NN+NNN interactions (and are extrapolated to the TL). 
The solid line corresponds to the Lindhard function describing the response of a non-interacting Fermi gas. \label{fig:NewFig}}
\end{center}
\end{figure}

Similarly to what 
was shown in Figs.~\ref{fig:AFDMC0.1resp} \& \ref{fig:AFDMC0.04resp}, one can extract the response function of neutron matter for other 
densities. We have carried out precisely such an extraction and show the results in Fig.~\ref{fig:NewFig}. 
These correspond to AFDMC calculations using NN+NNN interactions for the case where two periods
fit inside the box. They are compared to the free-gas result, which follows from the Lindhard 
function. Overall, we see that the microscopic results are roughly similar to the free-gas
results regardless of the density. At a more fine-grained level, we observe that the answer
to whether or not the microscopic response is higher or lower than the free-gas one depends
on the density. One could say that this behavior is similar to what is seen in Fig.~\ref{fig:homogeneous}, but such
a comparison is misleading for two reasons: a) there we were comparing AFDMC results to EDF results,
not to free-gas values, and b) our new results in Fig.~\ref{fig:NewFig} show the answer for the response, i.e.
at finite one-body potential strength. Thus, these responses cannot be simply extracted from  
the homogeneous-gas answers and constitute microscopic benchmarks.

\section{Summary \& Conclusion}

To summarize, in this article we have investigated the properties of periodically modulated neutron matter, using 
a combination of large-scale simulations and qualitative insights. We started from the non-interacting problem,
examining in detail finite-size effects: since 66 is the number of particles commonly used for homogeneous neutron matter, 
we studied the adjustments that need to be carried out in order to use that particle number for the inhomogeneous problem.
We then reported on our auxiliary-field diffusion Monte Carlo simulations, underlining the importance of optimizing the trial wave function by minimizing the VMC energy. This depended on a detailed understanding of the single-particle orbitals.
AFDMC allowed us to compute the ground-state energy of neutron matter at various densities, potential strengths, and periodicities of the potential. In particular we studied the inhomogeneous problem by increasing the strength of the potential starting from homogeneity.

We then examined several consequences of our \textit{ab initio} results. We first saw the impact that they have on energy density functionals. We used the response of neutron matter to constrain the isovector term while carefully disentangling the contributions of bulk and gradient terms. We found a density-dependent isovector term and provided our estimate for its
magnitude at each density. Next, we extracted the linear density-density static response function of neutron matter from AFDMC and EDF results at two different densities. This required a set of \textit{ab initio} results for each of the periodicities that we studied. We then compared and contrasted the response function of neutron matter to that of other systems. 
More than a proof-of-principle, this work provides detailed benchmarks that other \textit{ab initio} calculations can compare
to, or EDF approaches can use as input.

The authors are grateful to A. Bulgac, J. Carlson, S. Gandolfi, J. W. Holt, C. J. Pethick, S. Reddy, and A. Rios for many valuable discussions. They would also like to thank D. Lacroix and A. Pastore for sharing their Skyrme results as well 
as several insights. This work was supported by the Natural Sciences and Engineering Research Council (NSERC) of Canada and the Canada Foundation for Innovation (CFI). Computational resources were provided by SHARCNET and NERSC.

\end{document}